\documentclass{article}
\font\fthreei=cmti10 scaled\magstep1

\font\fthreebb=cmbx10 scaled\magstep2

\begin{document}

\centerline {\fthreebb ON KANT'S FIRST INSIGHT INTO THE}
\vskip 0.09 true cm
\centerline {\fthreebb PROBLEM OF SPACE DIMENSIONALITY }
\vskip 0.09 true cm
\centerline {\fthreebb AND ITS PHYSICAL FOUNDATIONS}

\vskip 1.2 cm
\centerline {\fthreei F. Caruso$^{1,2}$ \& R. Moreira Xavier$^1$}
\bigskip
\smallskip
\centerline {1. Centro Brasileiro de Pesquisas F\'{\i}sicas}
\centerline {Rua Dr. Xavier Sigaud 150, 22290-180, Rio de Janeiro, RJ, Brazil}

\medskip
\centerline {2. Brazilian Academy of Philosophy}
\centerline {Rua Riachuelo, 303, 20230-011, Rio de Janeiro, RJ, Brazil}

\vskip 3.0 cm

\centerline {\textbf{Abstract}}
\vskip 0.6 true cm

\noindent In this article it is shown that a careful analysis of Kant's \textit{Gedanken von der wahren Sch\"{a}tzung der lebendigen Kr\"{a}fte und Beurtheilung der Beweise} leads to a conclusion that does not match the usually accepted interpretation of Kant's reasoning in 1747, according to which the young Kant supposedly establishes a relationship between the tridimensionality of \textit{space} and Newton's law of universal gravitation. Indeed, it is argued that this text does not yield a satisfactory explanation of \textit{space} dimensionality, actually restricting itself to justify the tridimensionality of \textit{extension}.
\vfill\eject

\section{Introduction}
\bigskip

It was during the flourishing period of the mechanization of the World view, more precisely in 1747, that Kant tried to understand why space (\textit{Raum}) is tridimensional, in his first writing titled \textit{Gedanken von der wahren Sch\"{a}tzung der lebendigen Kr\"{a}fte und Beurtheilung der Beweise, deren sich Herr von Leibniz und andere Mechaniker in dieser Streitsache bedient haben, nebst einigen vorhergehenden Betrachtungen, welche die Kraft der K\"{o}rper \"{u}berhaupt betreffen} (\textit{GSK}).\footnote{Kant, 1747. Akademie-Ausgabe (AA): GSK, AA 01.} The text is divided in three sections, and only the first one will be object of study throughout this paper. It is largely accepted that the analysis carried on that first section establishes a clear framework for the discussion of space dimensionality as a problem in Physics, and represents its first physical solution.\footnote{Jammer, 1993; Brittan, 1978; Barrow, 1983; Barrow \& Tipler, 1986; Caruso \& Moreira Xavier, 1987.}  This contribution is generally summed up in the statement that \textit{the reason for the tridimensionality of space can be found in Newton's law of gravitation, according to which the force between two bodies decays with the square of the distance separating them}.\\

However, it will be seen throughout this paper that a more careful reading of \textit{GSK} leads us to conclude -- other than that is normally accepted -- that Kant's reasoning does not actually lead to a satisfactory determination of the \textit{space dimensionality}, but limits itself to justify the tridimensionality of \textit{extension} (\textit{Ausdehnung}). In any case, it can be argued that the original approach, conceived by Kant in his youth, strongly depends on the Newtonian concept of \textit{force}. More specifically, this means that Kant assumes \textit{force} as the \textit{causa efficiens} of the World changes. In this way, in this earliest text, he attempts to articulate Philosophy and Scientific knowledge trying to solve the \textit{vis viva} controversy. It is in such a framework that the problem of dimensionality is discussed.\\

Although the basic idea of somehow relating dimensionality to the Gravitation law was abandoned during the critical period of Kantian philosophy, Kant's attempt to determine space dimensionality from a physical law is, unquestionably, a milestone in the modern discussion of this essential attribute of space.\footnote{Barrow, 1983; Barrow \& Tipler, 1986; Caruso \& Moreira Xavier, 1987.} \\

In spite of the fact that this fruitful conjecture is part of a well known and discussed text, we feel that there are still some open questions which need to be clarified, namely:

\medskip

\begin{itemize}
\item[--] What are the bases of Kant's conjecture?
\item[--] Why does Kant ultimately limit himself, in his pre-critical phase, to explain the tridimensionality of {\it extension} rather than
that of {\it space}?
\item[--] Did Kant's pre-critical theories (mainly his space and time concepts) have any repercussion in the concepts developed by him in the critical period?\footnote{Baker, 1935.}
\end{itemize}

In this essay, Kant's first text is revisited, with just the scope to cast some light on aspects raised by the first two questions above.

\vskip 0.6 true cm
\section{The scenario where Kant's \textit{GSK} was written}
\bigskip

It is not our intention in this paper to discuss neither Kant's broad interest on Science\footnote{This issue is well discussed in the literature. See, for example, Britan, 1978; Falkenburg, 2000;  Friedman, 1998. Grillenzoni, 1998; Hoppe, 1969; Martin, 1965; Plaass, 1965; Waschkies, 1987.}  nor how he developed his pre-critical conception of space.\footnote{J.T. Baker, 1935.}  However, some general remarks concerning the philosophical and scientific frameworks where his \textit{GSK} was written are important to the comprehension of his specific attempt to deduce space dimensionality from experience.\\

In the plane of the history of ideas, Kant's first published work was inserted in different important and enthusiastic debates concerning: the nature of forces; the explanation of change; and the nature of space.\footnote{Garnett, 1939, pp.~89-118.}  Ultimately, one can say that this work, as well as the other pre-critical writings, is concerned with an examination of what Descartes, Leibniz and Newton had claimed about those points. It can be considered as a result of the hybrid confluence of critical analysis and acceptance of many of those competing claims.\\

First of all, Kant had to discuss the nature of the motion in order to face the problem of the dispute between Cartesians and Leibnizians about the proper measure of force. In \textit{GSK}, he chooses an intermediary position between the proposals of these two schools by maintaining that both quantities $mv$ and $mv^2$ could be conserved in different contexts.\footnote{Nevertheless, this question is not relevant so far as the problem of space dimensions is concerned.}  Secondly, he had to take position on the issue of the \textit{body-soul} problem. At that time, among the three possibilities -- occasionalism, pre-stablished harmony and \textit{influxus physicus} -- he takes position in favor to the last one.\footnote{Laywine, 1993, pp.~25-42.}  This means to admit that ``if creatures can act on one another, all change in creation is the effect of real interaction. Perceptions are the effect of the body's agency on the soul,\footnote{For the young Kant, the inner state of the soul can be understand as a \textit{status repraesentativus universi}.}  and voluntary motion is the effect of the soul's agency on the body''.\footnote{Laywine, \textit{ibid}.}  As will be seen in detail in the next Sections, Kant chooses Newton's Gravitational force to be this agent.\footnote{And, implicitly, the Newtonian law of action and reaction.}  Last but not least, the \textit{GSK} is not imune to the famous debate concerning the relative or absolute nature of space, which can be summarized in the Leibniz-Clarke polemics.\footnote{Alexander, 1956.}\\

In a nutshell, the young Kant was trying to provide a foundation for the metaphysics of nature, understanding that its task is to discover the inner force of things, the first causes of the law of motion and the ultimate constituents of matter.  In order to grasp the Newtonian influence at that point it is enough to remember what is written in the Preface of the first edition of Newton's \textit{Principia}: ``(...) for the whole burden of philosophy seems to consist in this -- from the phenomena of motions to investigate the forces of nature and then from these forces to demonstrate the other phenomena''.\footnote{We quote from Motte's translation of the \textit{Principia} revised by Cajori, reprinted by University of California Press, 1934. The latim text is: \textit{Omnis enim Philosophiae difficultas in eo versari videtur, ut a Phaenomenis motuum investigemus vires Naturae, deinde ab his viribus demonstremus phaenomena reliqua}.}\\

The hybrid way the young Kant treated all the aforementioned controversies allowed him to consider the problem of space dimensionality -- as being both a metaphysical and physical problem -- which is what we want to discuss from now on. \\

\vskip 0.6 true cm
\section{Kant and the Natural Philosophy of Space:
between Leibniz and Newton}
\bigskip

Probably under the influence of his professor Martin Knutzen, Kant developed, while at the university,\footnote{Grillenzoni, 1998, pp.~23-46.} a special interest in Physics and Mathematics, which evolved in such a way that his pre-critical writings essentially dealt with Physics, Cosmology\footnote{Falkenburg, 2000.} and the study of volcanoes. Knutzen also offers to his student the Wolffian metaphysical principles and shares with him many issues that are amalgamated in Kant's GSK, namely: the (im)possibility of the existence of other worlds and the body-soul problem.\footnote{Remember that Knutzen has published the \textit{Dissertatio metaphysica de aeternitate mundi impossibili} (which is known was read by Kant) and, in 1735, \textit{Commentatio philosophica de commercio mentis et corporis per influxum physicum explicando, ipsis illustrii Leibnitii principiis superstructura}.} Implicit in those problems, as already mentioned, there are influences from Leibniz, from Descartes and from Newton.\footnote{Newton's influence over Kant actually transcends the pre-critical period, in the sense that Kant tried, along all his works, to build metaphysics as science, in a way similar to the Newtonian system. \textit{Cf.} Mathieu, 1976.} \\

On one hand, the very title of Kant's first work, \textit{GKS}, shows clear traces of Leibniz's influence. On the other hand, Newtonian conceptions are the gist of his argument: the previous existence of substances -- able to interact through forces -- is crucial to the development of Kant's proof, as the title of the ninth paragraph of his text about \textit{The Living Forces} (\textit{GSK}) suggests. The \textit{Randtext} found in the beginning of the ninth paragraph states his intention to discuss the tridimensionality of space: ``(...) \textit{if the \underbar {substances} had no \underbar {force} whereby they can act outside themselves, there would be no extension, and consequently no space}''.\footnote{All English texts of Kant's quotations are from Handyside's translation of 1929. The reference to the original text is AA 01: 23. \textit{Randtext}.  ``(...) \textit{wenn die Substanzen keine Kraft h\"{a}tten au\ss{er} sich zu wirken, so w\"{u}rde keine Ausdehnung, auch kein Raum sein''}. The underlining is ours.}
\medskip

In fact, one of the first problems Kant was concerned to was related to the corporeal matter and to the interaction of physical substances. How to express this interaction in universal terms of cause and effect, and in which way is matter (the substance) able to alter the state of the soul by means of the force it possess in its motion, are issues about which he reflected. Furthermore, he accepts Leibniz' idea that the bodies have something (\textit{aliquid}) -- an inherent, essential force --, prior even to extension itself: ``\textit{[In rebus corporeis] est aliquid praeter extensionem, imo extensione prius, [alibi admonuimus]}.''\footnote{Leibniz, 1695, \textit{apud} Kant, GSK, AA 01: 17.22-23.}  Thus, the young Kant agrees with Leibniz that force precedes extension,\footnote{``The very substance of things consists in the force of action and passion'' (\textit{Die philosophischen Schriften von G.W. Leibniz}, ed. C.J. Gerhardt, Berlin, 1875-90, IV, p.~508.}  which is indeed supposed to be founded on the abstract concept of force. Since this concept is conceived by the human intellect, the net Cartesian distinction between \textit{res extensa} and \textit{res cogitans} fades away.\\

Let us now comment on Kant's concept of extension. Descartes, in his \textit{Principles}, trying to accommodate Aristotle's thesis that, since interpenetration of dimensions is impossible, there is no space apart from bodies, identifying ``space, or internal place'' with the ``corporeal substance contained in it''.\footnote{\textit{Apud} Des Chene, 1996, pp.~360-1.}  In addition, Descartes claimed that ``the extension in length, width, and depth that constitutes space is evidently the same as that which constitutes body''.\footnote{\textit{Idem}.}  The Cartesian concepts of extension and impenetrability are essential to his metaphysical project to reduce matter to Geometry.\footnote{Powers, 1991.}  Leibniz, on the other hand, defended the view that \textit{extension} was an attribute of \textit{substance}. Thus, concerning the concept of substance, Kant largely adhered to Leibniz's point of view. In other words, he accepted that extended objects occupy space not because of their extension, but by virtue of the dynamic qualities of impenetrability and resistance.\\

There is, however, one point of disagreement with Leibniz. This is when Kant makes the difference between \textit{vis activa} e \textit{vis motrix} in order to promote the triumph of \textit{influxus physicus} over the pre-stablished harmony supported by Leibniz.\footnote{Vuillemann, 1955, p.~233.}  \\

Concerning the nature of space, it is clear that Kant admits in the \textit{GSK} a relational space in a Leibnizian way, and not a Newtonian space conceived as a receptacle of bodies and phenomena. So, let us now briefly discuss why Kant's logical construction of the concept of space led him to keep away from Newtonian absolute space.\\

In all Kant's cognitive process developed in \textit{GSK}, forces play a fundamental role. In a certain sense, his viewpoint reminds us the stoic idea, of great impact during the last three centuries b.C.,\footnote{Sambursky, 1987.} that there exists a force -- which everything permeates --, due to the interaction of \textit{pneuma} and ponderable matter, and that this force creates a well-ordered \textit{continuum}, called \textit{space}.\\

In Kant's opinion, it is through these forces that connection (\textit{Verbindung}) among bodies can be established, from which the \textit{necessary order} to the existence of space is achieved.\footnote{It is worth to notice that the young Kant contribution to the theme treated in this paper is a rupture with the Aristotelian view of the issue -- both in its general realm (the cause of the space) and in its particular aspect (the cause of dimensionality) --, through the introduction of \textit{force} as the \textit{causa efficiens} of space, through the concept of \textit{order}. Although in a certain sense he is Aristotelic, considering the role played by the concept of \textit{substance} used in his discussion of space dimensionality, it should be stressed that, contrary to Aristotle -- in whose system force (\textit{dynamis}) leads to the \textit{rupture} of cosmic order -- Kant considers, in his first text, \textit{force} as a \textit{generator} of \textit{order}.} This can be seen from the following passage:

\begin{quotation}
\noindent ``\textit{It is easily proved that there would be no
space and no extension, if substances had no force whereby they can act
outside themselves. For without a force of this kind there is no
connection, without this connection no order, and without this order no
space.''}\footnote{GSK, AA 01: 23: 05-09. {\it ``Es ist leicht zu erweisen, da\ss\ kein Raum und keine Ausdehnung sein w\"{u}rden, wenn die Substanzen keine Kraft h\"{a}tten, au\ss{er} sich zu wirken. Denn ohne diese Kraft ist keine Verbindung, ohne diese keine Ordnung, und ohne diese endlich kein Raum.''}}
\end{quotation}

From this quotation we clearly see that without this \textit{force} (\textit{Kraft}), from which the substances are able to act externally, there would be no \textit{connection} between things, no order and therefore no \textit{space} (in this sequence). Therefore, a kind of \textit{dynamical connection} among substances is implicit here, from which one can safely infer that Kant is denying Leibniz's concept of \textit{monad}, since monads lack spatial extension and have no true causal relation with other monads. At this point, it is important to remember that ``Leibniz only advanced to his notion of space and time as relatives when he had redefined ultimate scientific objects as monads''.\footnote{Baker, \textit{op.~cit.}, p.~270.}  Thus, in his first writing, Kant is giving an original justification of the relative nature of space. \\

As Handyside well remarks in his Introduction to the English translation \textit{Inaugural dissertation and the early writings on space},\footnote{Cited as Kant's English translation, 1929 (1979).}  in this phase, ``Kant considers the space as a \textit{subsidiary phenomenon},\footnote{The emphasis is ours.}  which depends on the intelligible relations of these substances''. Such relations being expressed by force laws, it seems evident to us that, although Kant already accepts the core of the Newtonian scientific program, he (at least during this period) diverges from Newton in a crucial point of his system, namely the essence of space.\\

In fact, absolute space and time are, according to Koyr\'{e}, ``(...) \textit{r\'{e}alit\'{e}s que Newton acceptait sans h\'{e}siter -- puisqu'il pouvait les appuyer sur Dieu et les fonder en Dieu}''.\footnote{Koyr\'{e}, 1971.}  But for the young Kant, space \textit{is not} the divine \textit{sensorium}. On the contrary, it is an ideal substance-dependent construction, able to express and to emphasize particularly the role Reason plays. Naturally this is not the Cartesian identification of \textit{space}, \textit{quantity} and \textit{corporeal substance},\footnote{Mamiani, 1981.}  for here \textit{forces} cause \textit{extension}.\footnote{This pre-critical analysis puts Man (and not God) at the center of the discussion about space and its qualities.}  It is important to emphasize this point, since -- as we shall see below -- Kant's justification of tridimensionality depends strongly on \textit{this} concept of space, in opposition to the one he will adopt in his critical period.

\vskip 0.6 true cm
\section{Kant and the Tridimensionality}
\bigskip

Although Kant announces, in his first text, the intention to discuss the tridimensionality of space, one can argue whether it is (or not) his conception of space (in the pre-critical phase) which allowed him to lay only the basis for the tridimensionality of the extension.  Indeed, in the margin comment to the ninth paragraph, Kant states that ``\textit{the ground of the threefold dimension of space is still unknown.''}\footnote{GSK AA 01: 23, \textit{Randtext}. {\it ``Der Grund von der dreifachen Dimension des Raumes ist noch undekannt."}}
and, in the \textit{Randtext} of the next paragraph, he suggests a possible relation between the tridimensionality of space and the law of attraction between different bodies:
``\textit{It is  probable that the threefold dimension of space is due to the law according to which the forces in the substances act
upon one another.''}\footnote{  GSK AA 01: 24, \textit{Randtext}. ``\textit{Es ist wahrscheinlich, da\ss\ die dreifache Abmessung des Raumes von dem Gesetze herr\"{u}hre, nach welchem die Kr\"{a}fte der Substanzen in einander wirken.''}}\\

As Handyside comment in the \textit{Preface} to his translation of part of \textit{GSK},\footnote{\textit{Op.~cit.}, p.~xi.} it seems logical that being space a consequence of the connection of substances, 

\begin{quotation}
\noindent ``\textit{the special character of space as three-dimensional must, it would seem, find its explanation in the special character of law of the connections of substances. But what that is Kant declares himself unable to say (...) and in consequence he declares, logically enough, that space of other dimensions and other properties is possible.}''
\end{quotation}

\noindent as will be seen in the following. However, with no exception, throughout the text that follows the above quotation, corresponding to the very demonstration of the statement, Kant actually refers to the dimensionality of \textit{extension}.\\

Kant's reasoning, as seen in the previous Section, encompasses the following points: first, the idea that there exists a force inherent to the substances (that is to say, to the bodies), without which there would be no extension and no relation. Second, this force is necessary to establish the relations among the things, necessary to the order, and finally, without this order, space does not exist.\\

The fundamental role that the concept of \textit{force} -- the first essence of matter and of its extension -- plays in Kant's explanatory system is corroborated by the following quotation:

\begin{quotation}
\noindent  {\it ``Since everything which is to be found among the
qualities of a thing must be capable of being derived from that which
contains in itself the most complete ground of the thing itself, the
qualities of the extension, and subsequently their threefold dimension,
will be grounded in the qualities of the force which the substances possess
in respect of the things with which they are connected.''}\footnote{GSK, AA 01: 24.02-09. ``\textit{Weil alles, was unter den Eigenschaften eines Dinges vorkommt, von demjenigen mu\ss\ hergeleitet werden k\"{o}nnen, was den vollst\"{a}ndigen Grund von dem Dinge selber in sich enth\"{a}lt, so warden sich auch die Eingenschaften der Ausdehnung, mithin auch die dreifache Abmessung derselben auf die Eigenschaften der Kraft gr\"{u}nden, welche die Substanzen in Absicht auf die Dinge, mit denen sie verbunden sind, besitzen.''}}
\end{quotation}

From this quotation, independently of the subjacent force definition to be adopted (Cartesian, Leibnizian, or Newtonian), it is clear that this is the force through which the substances act upon one another, the one which is responsible for the collective relations which will, in Kant's view (in 1747), define space. As to the nature of this force, Kant states that

\begin{quotation}
\noindent {\it ``The force, whereby a substance acts in union
with others, cannot be thought apart from a determinate law which reveals
itself in the mode of its action. Since the character of these laws
according to which a whole collection of substances (that is, a space) is
measured, in other words, \underbar {the dimension of
extension},\footnote{The underlining is ours.} will
likewise be due to the laws according to which the substances by means of
their essential forces seek to unite themselves''.}\footnote{  GSK, AA 01: 24.09-18. ``\textit{Die Kraft, womit eine Substanz in der Vereinigung mit andern wirkt, kann nicht ohne ein gewisses Gesetz gedacht werden, welches sich in der Art seiner Wirkung hervorthut. Weil die Art des Gesetzes, nach welchem die Substanzen in einander wirken, auch die Art der Vereinigung und Zusammensetzung vieler derselben bestimmen mu\ss, so wird das Gesetz, nach welchem eine ganze Sammlung Substanzen (das ist ein Raum) abgemessen wird, oder die Dimension der Ausdehnung, von den Gesetzen herr\"{u}hren, nach welchen die Substanzen verm\"{o}ge ihrer wesentlichen Kr\"{a}fte sich zu vereinigen suchen.''}}
\end{quotation}

After these considerations, Kant stresses that the law of forces to which he refers is Newton's law of attraction which depends on the inverse of the square of the distance between to bodies. This choice seems to be due to the fact that Newton had shown that there is a principle connecting all the bodies of the Universe, which is expressed by an \textit{universal law}, the Gravitational law. At this point, however, he expresses himself with double caution, omitting whether the tridimensionality to which he refers is related to \textit{space} or to \textit{extension} (\textit{die dreifache Abmessung})  and concluding that \textit{it seems to result} (\textit{scheinet daher zu r\"{u}hren}) from the form of Newton's law of attraction, as the text below shows:

\begin{quotation}
\noindent {\it ``The threefold dimension seems to arise from the
fact that substances in the existing world so act upon one another that the
strenght of the action holds inversely as the square of the distances.''}\footnote{Margin comment to GSK, AA 01: 24. \textit{Randtext}.  ``\textit{Die dreifache Abmessung scheinet daher zu r\"{u}hren, weil die Substanzen in der existierenden Welt so in einander w\"{u}rken, da{\ss} die St\"{a}rke der Wirkung sich wie das Quadrat der Weiten umgekehrt verh\"{a}lt."}}
\end{quotation}

And then he adds:

\begin{quotation}
\noindent {``(...)\it that this law is arbitrary, and that God could have chosen another, for instance the inverse threefold relation; and lastly, that from a different law of extension with other properties and dimensions would have arisen.''}\footnote{GSK, AA 01: 24.26-30. ``(...)\textit{da\ss\ dieses Gesetz willk\"{u}rlich sei, und da Gott daf\"{u}r ein anderes, zum Exempel des umgekehrten  wirken, da\ss\ die dreifachen Verh\"{a}ltnisses, h\"{a}tte w\"{a}hlen k\"{o}nnen; da\ss\ endlich  St\"{a}rke der viertens aus einem andern Gesetze auch eine Ausdehnung  Wirkung sich  von andern Eigenschaften und Abmessungen geflossen w\"{a}re.''}}
\end{quotation}

\noindent after what, once again explicitly alluding to the dimension of \textit{extensions}, and not to space dimensionality (\textit{Ausdehnungen von andern Abmessungen}), he says that 

\begin{quotation}
\noindent {``\textit{If it is possible that there are extensions with other dimensions, it is also very probable that God has somewhere brought them into being; for His works have all the magnitude and manifoldness of which they are capable. Spaces of this kind, however, cannot stand in connection with those of a quite different constitution. Accordingly such spaces would not belong to our world, but must from separate worlds. Although in what precedes I have shown that it is possible that a number of worlds (in the metaphysical sense of the term) may exist together, in the considerations now before us we have, as it seems to me, the sole condition under which it would also be probable that a plurality of worlds actually exist.}''}\footnote{GSK, AA 01: 25.4-15. \textit{Wenn es m\"{o}glich ist, da\ss\ es Ausdehnungen von andern Abmessungen gebe, so ist es auch sehr wahrscheinlich, da\ss\ sie Gott wirklich irgendwo angebracht hat. Denn seine Werke haben alle die Gr\"{o}\ss{e} und Mannigfaltigkeit, die sie nur fassen k\"{o}nnen. R\"{a}ume von dieser Art k\"{o}nnten nun unm\"{o}glich mit solchen in Verbindung stehen, die von ganz anderm Wesen sind; daher w\"{u}rden dergleichen R\"{a}ume zu unserer Welt gar nicht geh\"{o}ren, sondern eigene Welten ausmachen m\"{u}ssen. In dem vorigen habe ich gezeigt, da{\ss} mehr Welten, im  metaphysischen Verstande genommen, zusammen existiren k\"{o}nnten; allein hier ist zugleich die Bedingung, die, wie mir deucht, die einzige ist, weswegen es auch wahrscheinlich w\"{a}re, da\ss\ viele Welten wirklich existiren}.}
\end{quotation}

So, we see that trying to give to the tridimensionality a basis on a kind of necessity, Kant now turns out around and explicitly notes the \textit{arbitrary} character of the laws of motion. Perhaps, this led him to realize that tridimensionality of space is \textit{contingent}. In addition, this also led him to speculate on the possible existence of different kinds of space and then on the existence of multiple Worlds,\footnote{Garnett, \textit{op.~cit.}, p.~102.} which was denied by Leibniz.  \\

It must be stressed that, except for the title, in no other part of this tenth paragraph does Kant explicitly use the word \textit{space} when referring to the tridimensionality, alluding to it only three times. The first and second, quoted above, only reinforce the idea of \textit{space} defined from \textit{physical substance}. The third, which also seems relevant to the theme treated here, is when Kant concludes his speculations by referring to \textit{space}, that is, to the various types of spaces, as objects of study of Geometry:

\begin{quotation}
\noindent {\it ``A science of all these possible kinds of space
would undoubtedly be the highest enterprise which a finite understanding
could undertake in the field of geometry.''}\footnote{GSK, AA 01: 24.31-33. ``\textit{Eine Wissenschaft von allen diesen m\"{o}glichen Raumesarten w\"{a}re unfehlbar die h\"{o}chste Geometrie, die ein endlicher Verstand unternehmen k\"{o}nnte.''}}
\end{quotation}

This fact can be considered as an indication of Kant's comprehension that the study of more generic spaces should precede the discussion of space dimensionality in Physics. Even though his result refers to the dimensions of \textit{extension}, Kant had to consider the possibility of the existence of spaces with a different number of dimensions, before any formal theory for these types of space.  It will be the Nineteenth Century discovery of non-Euclidean geometries that will give impulse to the discussion of these issues.\footnote{Jammer, 1993. Se also Hagar, 2008 and Kim, 2006.}   Anyhow, it seems to us that Kant, already in 1747, foresees that the road to the comprehension of the dimensionality of space should involve both Physics and Mathematics (through the generalization of Geometry to higher dimensional spaces), even if he could not, at that time, envision how his speculations would become important to the comprehension of space tridimensionality.\footnote{Whitrow, 1955.}  In any case, he sets the basis for modern discussions of this fascinating theme.\\

Looking forward in time, one can see that the road proposed by Kant leads to important developments. For example, one can quote the work of William Paley,\footnote{Payle, 1802.}  which can be considered the first attempt to shed light on the space dimensionality problem clearly from what nowadays is called Anthropic arguments.\footnote{Barrow \& Tipler, \textit{op.~cit.}}  In his work, Paley analyzes the consequences of changes in the form of Newton's gravitational law (the inverse square law) and of the stability of the solar system on human existence. Starting from a theological thesis, his speculations take into account a number of mathematical arguments which guarantee an anthropocentric design of the World, and rest upon the stability of the planetary orbits in our solar system and on a Newtonian mechanical \textit{Weltanschauung}. On the other hand, in the second half of the Nineteenth Century, the German mathematician Bernhard Riemann sustains that the metric structure of space depends on the matter distribution. Such a conjecture that matter ultimately determines some space properties is very similar to Kant's earlier ideas that physical (and geometrical) space depends on substance. In Riemann's word: ``The basis of metrical determination must be sought outside the manifold in the binding forces which act on it''.\footnote{``Es muss also entweder das dem Raume zu Grunde liegende Wirkliche eine discrete Mannigfaltigkeit bilden, oder der Grund der Massverh\"{a}ltnisse ausserhalb, in darauf wirkenden bindenden Kr\"{a}ften, gesucht warden''. Riemann, \textit{Collected Works},
p.~286. See also Riemann, 1873.}  Although these ideas did not resonate among the majority of the physicists and mathematicians in Riemann's time,  those questions were largely discussed by Einstein and are the \textit{fulcrum} of his General Theory of Relativity.\\

\vskip 0.6 true cm

\section{Final Considerations and Conclusions}
\bigskip

We have seen that Kant actually proposes a justification for the tridimensionality of \textit{extension} and not of \textit{space}, since he considers the latter as non-perceptible, as the product of an intellectual effort to seek to establish a kind of \textit{order} from the intelligible things. This \textit{space} appears as the object of study of Geometry and not of Physics. What really impresses the soul -- what is perceptible -- are the spatially extended objects, the \textit{matter} which causes effects on other \textit{substances}:
``(...) \textit{matter, by means of the force which it has
in its motion, changes that state of the soul whereby the soul represents
the world to itself.''}\footnote{GSK, AA 01: 21.22-24. ``(...) \textit{daher \"{a}ndert die Materie vermittelst ihrer Kraft, die sie in der Bewegung hat, den Zustand der Seele, wodurch sie sich die Welt vorstellet.''}}\\

It is possible to extract intelligible relations from substances starting from causal and universal laws of force like, particularly, Newton's law,\footnote{Such a causal relation was elaborated later by \"{U}berweg, 1882, \textit{apud} Jammer, 1993.} and the nature of space is dependent upon those substances. Garnett points out that Kant was arguing in circle, claiming that the force in bodies (or in the souls) lies in their outwardly directed acts.\footnote{``For this to occur, a body or a soul must have a position in space. Its position in space is, however, nothing but the resultant of its outwardly directed action upon other bodies and souls, and the actions of the others upon it. Kant begins by using position in space as a condition of the interaction of substances and went on to hold that such position is dependent upon the interaction''. Garnett, 1939, p.~95.} \\

From the physical point of view, a deeper comprehension of Kant's conjecture can only be reached by means of the \textit{field concept} in Physics. It is through the solution of Laplace-Poisson equation in $n$-dimensional Euclidean space that the relation between the exponent of the Newtonian potential and the dimensionality of space is straightforwardly established.\footnote{Ehrenfest, 1920; Tangherlini, 1963; Caruso \& Moreira Xavier, 1987.}  But it is only in the context of contemporary unified field theories that the problem of space dimensionality achieves the \textit{status} of a central problem in Physics: from this point of view one can realize the full meaning of Kant's contribution. 

Turning back to Kant's first text, we cannot ignore his intention to revisit this issue and the explicit warning at the end of the eleventh paragraph:

\begin{quotation}
\noindent {\it ``These thoughts may serve as an outline of an inquiry which I have in prospect. But I cannot deny that I expound them just as they have come into my mind, without that certainty which a more prolonged examination of them would have given. I am therefore ready to reject them, immediately a riper judment reveals to me their defects.''}\footnote{GSK, AA 01: 24.31-33. ``\textit{Diese Gedenken k\"{o}nnen der Entwurf zu einer Betrachtung sein, die ich mir vorbwhalte. Ich kann aber nicht leugnen, dass ich sie so mitteile, wie sie mir beifallen, ohne ihnen durch eine l\"{a}ngere Untersuchung ihre Gewissheit zu vershaften. Ich bin daher bereit, sie wieder zu verwerfen, so bald ein reiferes Urteil mir die Schw\"{a}che derselben aufdecken wird.''}}
\end{quotation}

As far as we know -- with the concordance of Brittan\footnote{Brittan, 1978.}  -- there is no other attempt by Kant aimed at giving a physical basis to the dimensionality issue. Indeed, afterwards Kant considered tridimensional space as an intuition \textit{a priori}. \\

It is known that Kant returned to this question, as certified by a manuscript included in the \textit{Opus Postumum}, but, ironically enough, there is an interruption in a key part of the text which makes it impossible for us to discover what the mature Kant would say about the problem of space dimensionality. We will therefore conclude this article with this reticent quotation by Kant:

\begin{quotation}
\noindent {\it ``The quality of space and time, for instance, that the
first has 3 dimensions, and the second only one, that the revolution is
ruled by the square of the distances are principles
that (...). [interruption]''}\footnote{OP, AA 22:13.29-31. ``\textit{Die Qvalit\"{a}t des Raumes u. der Zeit z. B. da\ss\ der erstere 3 Abmessungen die Zweyte nur Eine habe da\ss\ sich der Umkreis nach den Qvadraten der Entfernung richte sind Principien welche.}}
\end{quotation}
\vspace*{0.7cm}

\centerline {\bf Acknowledgment}
\bigskip
          The authors are in debt to an anonymous referee for pertinent comments and useful suggestions.

\vspace*{0.7cm}
\centerline {\bf {References}}
\bigskip

\noindent Alexander, H.G. (ed.): \textit{The Leibniz-Clarke Correspondence}. Manchester: Manchester University Press, 1956.\\

\noindent Baker, J.T.: Some pre-critical developments of Kant's theory of space and time, \textit{Philosophical Review} 44, 1935, 267-282.\\

\noindent Barrow, J.D.: Dimensionality, \textit{Philosophical Transactions of the Royal Society of London A} 310, 1983, 337.\\

\noindent Barrow, J.D. \& Tipler, F.J.: \textit{The Anthropic Cosmological Principle}. Oxford: Claredon Press, 1986.\\

\noindent Beiser, Frederick C.: Kant's intellectual development: 1746-1781, in Paul Guyer (ed.), \textit{The Cambridge Companion to Kant}. Cambridge: Cambridge University Press, 1992,
pp.~26-61.\\

\noindent Brittan, Jr., G.G.: \textit{Kant's Theory of Science}. Princeton: Princeton University Press, 1978.\\

\noindent Caruso, F. \& Moreira Xavier, R.: On the physical problem of spatial dimensions: an alternative procedure to stability arguments, \textit{Fundamenta Scientiae} 8, 1987, pp.~73-91.\\

\noindent De Chene, Dennis: \textit{Physiologia}. Natural Philosophy in Late Aristotelian and Cartesian Thought. Ithaca and London: Cornell University Press, 1996.\\

\noindent Ehrenfest, P.: Welche Rolle spielt die Dreidimensionalit\"{a}t des Raums in den Grundgesetzen der Physik. \textit{Annalen der Physik} 61, 1920, 440.\\

\noindent Falkenburg, Brigitte: \textit{Kant Kosmologie}. Die wissenschaftliche Revolution der Naturphilosophie im 18. Jahrhundert. Frankfurt am Main: Vittorio Klostermann, 2000.\\

\noindent Friedman, M.: \textit{Kant and the Exact Science}. Harvard: Harvard University Press, 1998. \\

\noindent Garnett Jr., Christopher Browne: \textit{The Kantian Philosophy of Space}. New York: Columbia University Press, 1939.\\

\noindent Grillenzoni, Paolo: \textit{Kant e la Scienza}. Vol. 1, 1747-1755. Milano: Vita e Pensiero, 1998.\\

\noindent Hagar, A.: Kant and non-Euclidean Geometry, \textit{Kant-Studien} 99, 2008, S. 80-98.\\

\noindent Hoppe, Hansgeorg. \textit{Kants Theorie der Physik}. Frankfurt am Main: Vittorio Klosterman, 1969.\\

\noindent Jammer, M.: \textit{Concepts of Space: the History of Theories of Space in Physics}, third edition. New York: Dover, 1993.\\

\noindent Kant, I.: \textit{Gedanken von der wahren Sch\"{a}tzung der lebendigen Kr\"{a}fte und Beurtheilung der Beweise, deren sich Herr von Leibniz und andere Mechaniker in dieser Streitsache bedient haben, nebst einigen vorhergehenden Betrachtungen, welche die Kraft der K\"{o}rper \"{u}berhaupt betreffen}, K\"{o}nigsberg, 1747; reprinted in: \textit{Kants gesammelte Schriften}. Herausgegeben von der K\"{o}niglich Preussischeen Akademie der Wissenschaften, Band I, Berlin 1902/1910. GSK, AA 01.\\

\noindent Kant, I.: \textit{Inaugural dissertation and the early writings on space}, translated by J. Handyside. Chicago: Open Court, 1929, reprinted by Hyperion Press, 1979.\\

\noindent Kant, I.: \textit{Opus Postumum -- passage des principes m\'{e}taphysiques de la science de la nature \`{a} la physique}. Translation, presentation and notes by F. Marty. Paris: Presses Universitaires de France, 1986, p.~131.\\

\noindent Kim, J.: Concepts and Intuitions in Kant's Philosophy of Geometry, \textit{Kant-Studien}, 97, 2006, S.~138-162.\\

\noindent Koyr\'{e}, A.: \textit{\'{E}tudes d'Histoire de la Pens\'{e}e Philosophique}. Paris: Gallimard, 1971, p.~269.\\

\noindent Laywine, Alison: \textit{Kant's early metaphysics and the origins of the critical philosophy}. Atascadero, California: Ridgeview Publishing Company, 1993.\\

\noindent Leibniz: \textit{Acta Eruditorum}, Leipzig, 1695.\\

\noindent Mamiani, M.: \textit{Teorie dello Spazio da Descartes a Newton}. Milano: Franco Angeli Editori, second edition, 1981.\\

\noindent Martin, Gottfried: \textit{Kant's Metaphysics and Theory of Science}. Manchester: Manchester University Press, 1955.\\

\noindent Mathieu, V.: Kant. In Centro di Studi Filosofici di Gallarate: \textit{Dizionario dei Filosofi}. Firenze: G.C. Sansoni, 1976, pp. 646-58.\\

\noindent Nuzzo, Angelica: \textit{Ideal Embodiment}. Kant's Theory of Sensibility. Bloomington and Indianapolis: Indiana University Press, 2008.\\

\noindent Paley, William: \textit{Natural Theology}, 1802, reprinted in \textit{The Works of William Paley}, 7 volumes, edited by R. Lynam, London, 1825. A recent book's reprint is available from Oxford University Press, 2006.\\

\noindent Plaass, Peter: \textit{Theorie der Naturwissenschaft}. G\"{o}ttingen: Vandenhoeck \& Rupprecht, 1965.\\

\noindent Polonoff, Irving: \textit{Force, Cosmos, Monads and Other Themes of Kant's Early Thought}. Bonn: Bouvier Verlag, 1973.\\

\noindent Powers, Jonathan: \textit{Philosophy and the News Physics}. London and New York: Routledge, 1991.\\

\noindent Riemann, B.: On the Hypotheses which lie at the Bases of Geometry (1868)
Translated by William Kingdon Clifford.  \textit{Nature} 8, 1873, n.~183, pp.~14-17.\\

\noindent Sambursky, S.: \textit{The Physical World of Late Antiquity}. London: Routledge \& Kegan Paul, 1987.\\


\noindent Tangherlini, F.R.: Schwarzschild field in $n$ dimensions and the dimensionality of space problem, \textit{Il Nuovo Cimento} 27 (1963) 636.\\

\noindent \"{U}berweg, F.: \textit{System der Logik und Geschichte der logischen Lehren}, fifth ed., Bonn: Adolph Marcus, 1882, S.~113.\\

\noindent Vuillemin, J.: \textit{Physique et M\'{e}taphysique Kantiennes}. Paris: Presses Universitaires de France, 1955. \\

\noindent Waschkies, Hans-Joachim: \textit{Physik und Physicotheologie dei jungen Kant}. Die Vorgeschichte seiner Allgemeinen Naturgeschichte und Theorie des Himmels. Amsterdam: Verlag B.R. Gr\"{u}ner, 1987.\\


\noindent Whitrow, G.J.: Why Physical Space has Three Dimensions?, \textit{British Journal for Philosophy of Science} 6, 1955, pp.~13-31.\\

\end{document}